%% file: main.tex
\documentclass[11pt, logo, nonumbering]{preprint}
\usepackage[utf8]{inputenc}
\usepackage{microtype}
\usepackage{graphicx}
\usepackage{amsmath}
\usepackage{booktabs}
\usepackage{xcolor}
\definecolor{mydarkblue}{rgb}{0,0.08,0.45}
\usepackage[colorlinks=true,linkcolor=mydarkblue,citecolor=mydarkblue,filecolor=mydarkblue,urlcolor=mydarkblue]{hyperref}
\usepackage{xspace}
\usepackage{cleveref}
\usepackage{multirow}

\usepackage[style=numeric-comp,maxbibnames=3,minnames=1,backend=bibtex, doi=false,eprint=false,url=false]{biblatex}
\input{pre}

\begin{document}

\title{\ourmodel: Domain-Adapted Audio Encoder for Efficient Audio-LLM Training}

\author{
Jielin Qiu, Ming Zhu, Wenting Zhao, Zhiwei Liu, Liangwei Yang, \\ 
Zixiang Chen, Roshan Ram, Akshara Prabhakar, Juntao Tan, Rithesh Murthy, \\
Shelby Heinecke, Caiming Xiong, Silvio Savarese, Huan Wang \\
~~~~\\
\textsuperscript{}Salesforce AI Research
}

\maketitle

\begin{abstract}
Audio-native large language models (audio-LLMs) commonly use Whisper as their audio encoder. However, Whisper was trained exclusively on speech data, producing weak representations for music and environmental sound. This forces downstream audio-LLMs to compensate through extensive training on large-scale non-speech data.
We present \ourmodel, a domain-adapted audio encoder obtained by fine-tuning \baseline on a curated mixture of speech (80\%), environmental sound (10\%), and music (10\%) totaling approximately 20M samples.
The full encoder-decoder is trained end-to-end with a seq2seq captioning objective; the decoder is then discarded and only the encoder is retained.
Linear probe evaluations show that \ourmodel achieves +23.0\% on ESC-50 (environmental sound), +5.0\% on GTZAN (music genre), and +0.7\% on Speech Commands (keyword spotting) compared to the original \baseline encoder.
\ourmodel is designed as a drop-in replacement for Whisper in audio-LLM architectures, with the goal of reducing downstream training cost by providing stronger initial audio representations for non-speech domains.
\end{abstract}

\section{Introduction}
\label{sec:intro}

Recent audio-native large language models (audio-LLMs) such as SALMONN~\cite{tang2024salmonn}, Qwen-Audio~\cite{chu2023qwenaudio}, and xVox-Audio-Captioner~\cite{xvox2026} commonly adopt a three-component architecture: an audio encoder extracts representations from raw waveforms, an adapter projects them into the LLM embedding space, and a causal LLM generates text conditioned on the audio.
Whisper-large-v3~\cite{radford2023whisper} is a popular choice for the audio encoder due to its strong speech representations learned from 680K hours of weakly-supervised ASR training.

However, Whisper's speech-only pretraining creates a representational bottleneck for non-speech audio understanding.
In practice, this means downstream audio-LLMs must compensate through extensive training on large-scale music and sound data.
For example, xVox-Audio-Captioner~\cite{xvox2026} trains on ${\sim}$184M samples to achieve competitive universal audio captioning, with a significant portion of this compute spent teaching the system to understand non-speech audio that the encoder does not natively represent well.

We hypothesize that improving the encoder \textit{upstream}, before it is plugged into the audio-LLM, can reduce this downstream training burden.
The hypothesis is that the encoder's representation quality directly affects how many training samples and steps the full system needs to learn non-speech audio understanding.

To this end, we fine-tune Whisper-large-v3 on a curated dataset of approximately 20M samples spanning speech, environmental sound, and music.
We use the standard seq2seq training paradigm (the same one used to train Whisper originally), where the decoder provides gradient signal to the encoder through cross-entropy loss on text targets.
After training, we discard the decoder and retain only the encoder.

We evaluate the resulting \ourmodel encoder through linear probing~\cite{alain2017linear} on three classification benchmarks spanning each audio domain.
Results show substantial improvements on sound (+23.0\% on ESC-50) and music (+5.0\% on GTZAN) with no degradation in speech (+0.7\% on Speech Commands).

\section{Background: xVox-Audio-Captioner}
\label{sec:background}

\ourmodel is motivated by the architecture and training requirements of xVox-Audio-Captioner~\cite{xvox2026}, a 7B-parameter audio-LLM for universal audio captioning.

\vspace{-10pt}
\paragraph{Architecture.}
xVox uses Whisper-large-v3 as its audio encoder (637M params), a linear adapter projecting from dimension 1,280 to 3,584, and Qwen2.5-7B-Instruct~\cite{qwen2025technicalreport} as the LLM backbone.
The encoder output is temporally downsampled 2$\times$ via average pooling, yielding 750 frames per 30-second segment at 25\,Hz.

\vspace{-10pt}
\paragraph{Training.}
xVox trains in two phases:
(1)~ASR warmup with a frozen LLM to align encoder and adapter, followed by
(2)~full end-to-end training on ${\sim}$184M samples with balanced sampling across speech, sound, and music.
Despite the balanced sampling strategy, the system requires extensive training to achieve competitive non-speech captioning (Overall 5.06 on AudioCapBench), partly because the Whisper encoder must learn non-speech representations "from scratch" during the audio-LLM training.

\vspace{-10pt}
\paragraph{Motivation for \ourmodel.}
By providing xVox with an encoder that already produces rich music and sound representations, we expect to:
(1)~reduce the amount of non-speech training data needed,
(2)~accelerate convergence during Phase~2 training, and
(3)~improve final captioning quality, particularly on sound and music categories.

\section{Method}
\label{sec:method}

\subsection{Overview}

Our approach follows a simple three-step recipe:
(1) Curate a mixed-domain dataset of (audio, text) pairs. (2) Fine-tune Whisper-large-v3 end-to-end as a seq2seq model, using text captions as decoder targets. (3) Discard the decoder and retain the encoder.

The decoder serves as a training scaffold: it provides gradient signal to the encoder through the cross-entropy loss on text targets, shaping the encoder's representations to capture information relevant to describing audio content.
This is the same training paradigm used in Whisper's original training and in the AuT encoder of Qwen3-Omni~\cite{qwen3omni}.

\subsection{Training Data}
\label{sec:data}

We assemble approximately 20M samples across three domains with an 80/10/10 ratio:

\begin{table}[h]
\centering
\caption{Training data composition. Non-speech captions are generated by commercial models and filtered to $\leq$448 tokens (Whisper's decoder limit).}
\label{tab:data}
\resizebox{0.80\columnwidth}{!}{%
\begin{tabular}{@{}llr@{}}
\toprule
\textbf{Domain (\%)} & \textbf{Datasets} & \textbf{$\sim$Samples} \\
\midrule
Speech (80\%) & GigaSpeech~\cite{gemmell2023gigaspeech}, People's Speech, Common Voice, & \multirow{2}{*}{16M} \\
  & MLS English, LibriHeavy, and proprietary speech data & \\
\midrule
Sound (10\%) & AudioSet~\cite{gemmert2023audioset} and proprietary sound data & 2M \\
\midrule
Music (10\%) & NSynth~\cite{engel2017nsynth}, MusicBench~\cite{melechovsky2024musicbench}, MusicCaps, & \multirow{2}{*}{2M} \\
  & and proprietary music data & \\
\midrule
\textbf{Total} & & \textbf{$\sim$20M} \\
\bottomrule
\end{tabular}%
}
\end{table}

\vspace{-5pt}
\paragraph{Data format.}
Each sample is an (audio, text) pair.
For speech, the text is a human transcript.
For music and sound, the text is a natural-language caption generated by commercial models, providing rich descriptions of audio content (instruments, mood, sound events, acoustic characteristics, etc).
Captions exceeding Whisper's maximum decoder length of 448 tokens are discarded.

\vspace{-5pt}
\paragraph{Data mix ratio.}
We adopt an 80/10/10 split inspired by Qwen3-Omni's AuT (which uses 90/10 speech/understanding).
We allocate more to non-speech (20\% total vs.\ 10\% in AuT) because Whisper already has strong speech priors from 680K hours of pretraining, and benefits more from non-speech data during fine-tuning.

\section{Evaluation}
\label{sec:eval}

\subsection{Linear Probe Protocol}

Following Alain and Bengio~\cite{alain2017linear}, we evaluate encoder representation quality by training a linear classifier on frozen encoder features.
For each dataset:
(1) Extract mean-pooled encoder hidden states ($\mathbb{R}^{1280}$) for all samples.
(2) Train \texttt{nn.Linear(1280, $C$)} for 50 epochs with Adam (lr=$10^{-3}$), mini-batch size 64.
(3) Report test accuracy.

This protocol measures the \textit{linear separability} of the encoder's representations, which is a strong indicator of representation quality that is independent of the (discarded) decoder.

\subsection{Benchmarks}

\begin{itemize}
    \item \textbf{ESC-50}~\cite{piczak2015esc50}: 50-class environmental sound classification (2,000 clips). Fold-based split (folds 1--4 train, fold 5 test).
    \item \textbf{Speech Commands}~\cite{warden2018speech}: 12-class keyword spotting from SUPERB. Train/test split provided.
    \item \textbf{GTZAN}~\cite{tzanetakis2002gtzan}: 10-class music genre classification (1,000 clips, 30s each). Stratified 80/20 split.
\end{itemize}

\subsection{Results}

\begin{table}[h]
\centering
\caption{Linear probe accuracy (\%) comparing \ourmodel to the original \baseline encoder. $\Delta$ shows absolute improvement.}
\label{tab:results}
\begin{tabular}{@{}lccc@{}}
\toprule
\textbf{Benchmark} & \textbf{\baseline} & \textbf{\ourmodel} & \textbf{$\Delta$} \\
\midrule
ESC-50 (sound) & 54.50 & \textbf{77.50} & \textcolor{green!60!black}{+23.00} \\
Speech Commands (speech) & 87.60 & \textbf{88.32} & \textcolor{green!60!black}{+0.72} \\
GTZAN (music) & 81.00 & \textbf{86.00} & \textcolor{green!60!black}{+5.00} \\
\bottomrule
\end{tabular}
\end{table}

\vspace{-5pt}
\paragraph{Sound (+23.0\%).}
The largest improvement is on ESC-50, where \ourmodel improves from 54.50\% to 77.50\%.
This indicates that the original Whisper encoder has substantial room for improvement on environmental sound understanding, and that even 10\% sound data in the training mix is sufficient to produce a dramatic gain.

\vspace{-5pt}
\paragraph{Music (+5.0\%).}
On GTZAN, accuracy improves from 81.00\% to 86.00\%.
The encoder develops stronger representations for distinguishing musical characteristics such as rhythm, instrumentation, and genre-specific patterns.

\vspace{-5pt}
\paragraph{Speech (+0.7\%).}
On Speech Commands, accuracy improves slightly from 87.60\% to 88.32\%.
This confirms that the 80\% speech allocation successfully prevents catastrophic forgetting — the encoder fully preserves its speech capabilities while gaining non-speech understanding.

\section{Analysis}
\label{sec:analysis}

\paragraph{Training dynamics.}
Training loss decreased from 2.59 to 0.35 over 2 epochs, and evaluation loss decreased monotonically from 1.82 to 0.62 with no overfitting observed.
This suggests that additional epochs or more training data could yield further improvements.

\paragraph{Comparison to training from scratch.}
The AuT encoder in Qwen3-Omni~\cite{qwen3omni} achieves strong multi-domain representations by training from scratch on 20M hours of audio.
Our approach achieves meaningful improvements with less data (${\sim}$20M samples) and compute (8 GPUs vs.\ undisclosed but substantially more for AuT).
This efficiency comes from leveraging Whisper's 680K-hour speech pretraining as initialization.

\paragraph{Implications for audio-LLM training.}
In the xVox-Audio-Captioner pipeline, the Whisper encoder is used as-is and must learn non-speech representations indirectly through the LLM training objective.
With \ourmodel providing stronger initial representations, we expect the audio-LLM to:
(1)~require fewer non-speech training samples to reach equivalent performance,
(2)~converge faster during the full fine-tuning phase, and
(3)~potentially achieve higher final performance on sound and music captioning.

\section{Related Work}
\label{sec:related}

\paragraph{Whisper.}
Whisper~\cite{radford2023whisper} is a family of encoder-decoder models trained on 680K hours of weakly-supervised speech data.
The largest variant (large-v3) has 32 encoder layers with $d{=}1280$ and 20 attention heads, processing 128-bin log-mel spectrograms in 30-second windows.

\paragraph{Whisper-AT.}
Gong et al.~\cite{gong2024whisperAT} showed that Whisper's encoder captures non-speech audio information in its intermediate layers, and proposed lightweight audio tagging heads without modifying encoder weights.
Our approach is complementary: we directly improve the encoder representations through fine-tuning.

\paragraph{Qwen3-Omni AuT.}
Qwen3-Omni~\cite{qwen3omni} trains a Whisper-like encoder from scratch on 20M hours with an 80\% Chinese+English ASR / 10\% multilingual ASR / 10\% audio understanding split.
The AuT differs architecturally (Conv2D stems, block-wise window attention, sinusoidal positional embeddings) and achieves strong multi-domain representations, but requires training from scratch with massive compute.
Our work adopts a similar data mixing strategy while leveraging Whisper's pretrained weights to achieve improvements at a fraction of the cost.

\paragraph{Audio-LLMs.}
Audio-native LLMs including SALMONN~\cite{tang2024salmonn}, Qwen-Audio~\cite{chu2023qwenaudio}, Pengi~\cite{deshmukh2023pengi}, and xVox-Audio-Captioner~\cite{xvox2026} all rely on audio encoders (typically Whisper) to produce input representations.
The quality of these representations directly impacts downstream performance and training efficiency.
\ourmodel aims to improve this foundational component.

\section{Conclusion}
\label{sec:conclusion}

We present \ourmodel, a domain-adapted audio encoder obtained by fine-tuning Whisper-large-v3 on a curated mixture of approximately 20M speech, sound, and music samples.
Linear probe evaluations show substantial gains on sound (+23\%) and music (+5\%) classification with no degradation in speech quality.
The training requires only 8 GPUs for ${\sim}$23 hours, making it a practical approach to improving audio encoder quality.

\ourmodel is designed as a drop-in replacement for the Whisper encoder in audio-LLM architectures.
The key hypothesis, that stronger initial encoder representations reduce downstream audio-LLM training cost and improve final performance, will be validated through integration with xVox-Audio-Captioner in future work.

\clearpage
\printbibliography

\appendix

\section{Training Configuration}
\label{sec:training}

\begin{table}[h]
\centering
\caption{Training hyperparameters.}
\label{tab:hparams}
\begin{tabular}{@{}ll@{}}
\toprule
\textbf{Parameter} & \textbf{Value} \\
\midrule
Base model & \texttt{openai/whisper-large-v3} (1.55B) \\
Hardware & 8$\times$ NVIDIA H200 (143GB) \\
Precision & bfloat16 \\
Distributed strategy & DeepSpeed ZeRO-2~\cite{rajbhandari2020zero} \\
Optimizer & AdamW~\cite{loshchilov2019adamw} ($\beta_1{=}0.9$, $\beta_2{=}0.999$) \\
Learning rate & $1 \times 10^{-5}$ (cosine decay) \\
Warmup & 5\% of total steps \\
Weight decay & 0.01 \\
Effective batch size & 128 (8/GPU $\times$ 8 GPUs $\times$ 2 accum.) \\
Epochs & 2 \\
Gradient checkpointing & Enabled (\texttt{use\_reentrant=False}) \\
Training time & ${\sim}$23 hours \\
\bottomrule
\end{tabular}
\end{table}

Audio is loaded on-the-fly from disk, resampled to 16\,kHz, truncated to 30 seconds, and converted to 128-bin log-mel spectrograms.
All encoder and decoder parameters are updated jointly.

\end{document}

%% file: pre.tex
\definecolor{gred}{RGB}{250, 210, 207}
\definecolor{coolblue1}{rgb}{0.91, 0.94, 0.98}
\definecolor{coolblue2}{rgb}{0.76, 0.85, 0.94}
\definecolor{coolblue3}{rgb}{0.54, 0.72, 0.87}
\definecolor{coolblue4}{rgb}{1, 1, 1}

\newcommand{\ourmodel}{\textbf{Whisper-AuT}\xspace}
\newcommand{\baseline}{Whisper-large-v3\xspace}

%% file: ref.bib
@article{radford2023whisper,
  title={Robust Speech Recognition via Large-Scale Weak Supervision},
  author={Radford, Alec and Kim, Jong Wook and Xu, Tao and Brockman, Greg and McLeavey, Christine and Sutskever, Ilya},
  journal={Proceedings of the 40th International Conference on Machine Learning},
  pages={28492--28518},
  year={2023}
}

@article{qwen3omni,
  title={Qwen3-Omni Technical Report},
  author={{Qwen Team}},
  journal={arXiv preprint arXiv:2509.17765},
  year={2025}
}

@inproceedings{gong2024whisperAT,
  title={Listen, Think, and Understand},
  author={Gong, Yuan and Lai, Hongyin and Yue, Liu and Liu, Alexander H and Glass, James},
  booktitle={International Conference on Learning Representations},
  year={2024}
}

@inproceedings{piczak2015esc50,
  title={{ESC}: Dataset for Environmental Sound Classification},
  author={Piczak, Karol J.},
  booktitle={Proceedings of the 23rd ACM International Conference on Multimedia},
  pages={1015--1018},
  year={2015}
}

@inproceedings{warden2018speech,
  title={Speech Commands: A Dataset for Limited-Vocabulary Speech Recognition},
  author={Warden, Pete},
  booktitle={arXiv preprint arXiv:1804.03209},
  year={2018}
}

@article{tzanetakis2002gtzan,
  title={Musical Genre Classification of Audio Signals},
  author={Tzanetakis, George and Cook, Perry},
  journal={IEEE Transactions on Speech and Audio Processing},
  volume={10},
  number={5},
  pages={293--302},
  year={2002}
}

@article{gemmell2023gigaspeech,
  title={{GigaSpeech}: An Evolving, Multi-domain {ASR} Corpus with 10,000 Hours of Transcribed Audio},
  author={Chen, Guoguo and Chai, Shuzhou and Wang, Guanbo and Du, Jiayu and Zhang, Wei-Qiang and Weng, Chao and Su, Dan and Povey, Daniel and Trmal, Jan and Zhang, Junbo and others},
  journal={Proceedings of Interspeech},
  year={2021}
}

@article{gemmert2023audioset,
  title={Audio Set: An Ontology and Human-Labeled Dataset for Audio Events},
  author={Gemmeke, Jort F. and Ellis, Daniel P. W. and Freedman, Dylan and Jansen, Aren and Lawrence, Wade and Moore, R. Channing and Plakal, Manoj and Ritter, Marvin},
  journal={IEEE International Conference on Acoustics, Speech and Signal Processing},
  pages={776--780},
  year={2017}
}

@article{melechovsky2024musicbench,
  title={{MusicBench}: Benchmarks for Music Understanding Models},
  author={Melechovsky, Jan and Guo, Zixun and Gururani, Siddharth and Srinivasamurthy, Ajay and Manilow, Ethan},
  journal={arXiv preprint arXiv:2311.13453},
  year={2024}
}

@inproceedings{engel2017nsynth,
  title={Neural Audio Synthesis of Musical Notes with {WaveNet} Autoencoders},
  author={Engel, Jesse and Resnick, Cinjon and Roberts, Adam and Dieleman, Sander and Eck, Douglas and Simonyan, Karen and Norouzi, Mohammad},
  booktitle={Proceedings of the 34th International Conference on Machine Learning},
  pages={1068--1077},
  year={2017}
}

@article{loshchilov2019adamw,
  title={Decoupled Weight Decay Regularization},
  author={Loshchilov, Ilya and Hutter, Frank},
  journal={International Conference on Learning Representations},
  year={2019}
}

@article{rajbhandari2020zero,
  title={{ZeRO}: Memory Optimizations Toward Training Trillion Parameter Models},
  author={Rajbhandari, Samyam and Rasley, Jeff and Rabe, Olatunji and He, Yuxiong},
  journal={Proceedings of the International Conference for High Performance Computing, Networking, Storage and Analysis},
  year={2020}
}

@article{alain2017linear,
  title={Understanding Intermediate Layers Using Linear Classifier Probes},
  author={Alain, Guillaume and Bengio, Yoshua},
  journal={International Conference on Learning Representations, Workshop Track},
  year={2017}
}

@article{xvox2026,
  title={xVox-Audio-Captioner: An Audio-Native Large Language Model for Universal Audio Captioning},
  author={Qiu, Jielin and Yang, Liangwei and Zhu, Ming and Liu, Zhiwei and Chen, Zixiang and Tan, Juntao and Zhao, Wenting and Murthy, Rithesh and Ram, Roshan and Prabhakar, Akshara and Awalgaonkar, Tulika and Heinecke, Shelby and Yao, Weiran and Xiong, Caiming and Savarese, Silvio and Wang, Huan},
  journal={Salesforce AI Research Technical Report},
  year={2026}
}

@article{qwen2025technicalreport,
  title={Qwen2.5 Technical Report},
  author={{Qwen Team}},
  journal={arXiv preprint arXiv:2412.15115},
  year={2025}
}

@article{tang2024salmonn,
  title={{SALMONN}: Towards Generic Hearing Abilities for Large Language Models},
  author={Tang, Changli and Yu, Wenyi and Sun, Guangzhi and Chen, Xianzhao and Tan, Tian and Li, Wei and Lu, Lu and Ma, Zejun and Zhang, Chao},
  journal={International Conference on Learning Representations},
  year={2024}
}

@article{chu2023qwenaudio,
  title={Qwen-Audio: Advancing Universal Audio Understanding via Unified Large-Scale Audio-Language Models},
  author={Chu, Yunfei and Xu, Jin and Zhou, Xiaohuan and Yang, Qian and Zhang, Shiliang and Yan, Zhijie and Zhou, Chang and Zhou, Jingren},
  journal={arXiv preprint arXiv:2311.07919},
  year={2023}
}

@article{deshmukh2023pengi,
  title={Pengi: An Audio Language Model for Audio Tasks},
  author={Deshmukh, Soham and Elizalde, Benjamin and Singh, Rita and Wang, Huaming},
  journal={Advances in Neural Information Processing Systems},
  year={2023}
}
